\documentclass[preprint,amssymb,amsmath,aip,jcp]{revtex4-1}

\usepackage[utf8]{inputenc}
\DeclareUnicodeCharacter{00A0}{~}
\usepackage[T1]{fontenc}
\usepackage[english]{babel}
\usepackage{graphicx}

\usepackage{amsmath}
\usepackage[makeroom]{cancel}
\usepackage{braket}
\usepackage{MnSymbol}

\usepackage{xcolor}

\usepackage{fullpage}
\usepackage{multirow}

\newcommand{\Ai}{\text{Ai}}
\newcommand{\rv}{\mathbf{r}}
\newcommand{\dr}{d\mathbf{r}}

\begin{document}

\title{Self-consistent assessment of Englert-Schwinger model on atomic properties}
\author{Jouko Lehtom\"aki}
\email{jouko.lehtomaki@aalto.fi}
\author{Olga Lopez-Acevedo}
\email{olga.lopez.acevedo@aalto.fi}
\altaffiliation{Universidad de Medell\'in, Colombia}
\affiliation{COMP Centre of Excellence, Department of Applied Physics, Aalto University, P.O. Box 11100, 00076 Aalto, Finland}
\begin{abstract}
Our manuscript investigates a self-consistent solution of the statistical atom model proposed by Berthold-Georg Englert and Julian Schwinger (the ES model) and benchmarks it against atomic Kohn-Sham and two orbital-free models of the Thomas-Fermi-Dirac (TFD)-$\lambda$vW family. Results show that the ES model generally offers the same accuracy as the well-known TFD-$\frac{1}{5}$vW model; however, the ES model corrects the failure in Pauli potential near-nucleus region. 
We also point to the inability of describing low-$Z$ atoms as the foremost concern in improving the present model.

\end{abstract}
\maketitle

\section{Introduction}
Orbital-free density functional theory (OFDFT) is a lucrative path to linear scaling DFT methods. The modern OFDFT theory is mostly investigated in the setting of non-local or generalized gradient approximation (GGA)-style kinetic functionals. Potential functionals, such as those used in Berthold-Georg Englert and Julian Schwinger’s statistical atom model \cite{statistical_atom_1,statistical_atom_2,statistical_atom_3}, have fallen out of favor.  Yet, it has been shown that at least GGA-style orbital-free functionals usually suffer from a theoretical flaw: the atomic nucleus is not well described, as evidenced by the singular Pauli potential \cite{pauli_potential_properties,pauli_potential_functionals,ofdft_issues}. The Pauli potential describes a positive effective repulsion arising from the Pauli exclusion principle, and its violation implies a fundamentally wrong solution of the electronic problem. Potential functionals allow linear-scaling DFT to sidestep this flaw and our intention here is to quantify the potential functionals’ self-consistent accuracy by benchmarking Englert and Schwinger’s model \cite{statistical_atom_2,statistical_atom_3} to more well-known OFDFT models.

A flurry of recent activity uses potential as a variable instead of density. Much of the work is focused especially on the Pauli potentials’ role and especially in atoms \cite{pauli_potential_differential_equation,pauli_potential_functionals,Finzel2016}. The potential variable has even been combined with machine learning\cite{potential_machine_learning}, which also has had some applications for the search of kinetic energy functionals \cite{ofdft_neural_networks,ofdft_machine_learning_functionals}. 

Pseudopotentials \cite{ofdft_pseudopotential,ofdft_issues} have been useful in conventional approaches to extended systems with OFDFT, where an active effort to improve local pseudopotentials' transferability is ongoing \cite{LPP_EA,LPP_OEP}. While this solution works remarkably well \cite{ofdft_pseudopotential}, a few questions remain for systematic improvement. In general, how much of the overall accuracy is because of the better ion-energy interaction description or the better electronic kinetic energy functional? Even more importantly, in the case of fitted pseudopotentials, when is a pseudopotential overfitted compared to the kinetic energy functional? One approach is to investigate the full potential solution to OFDFT equations, as we did by using the projector-augmented-wave method\cite{ofdft_paw}, but one of the problems here is the near-nucleus singularity of the Pauli potential, which brings the solution’s plausibility into question.

The problem near the atomic nucleus is not a new one, and it motivated the development of the statistical atom model \cite{statistical_atom_1}. Englert and Schwinger’s approach, however, is based on the effective potential instead of the density, which allows natural energy scale separation of the problematic core electrons from the bulk. In particular, we are interested in the model proposed in two papers \cite{statistical_atom_2,statistical_atom_3} and later refined in Englerts’ book, {\it Semiclassical Theory of Atoms} \cite{semiclassical_atom}.

The perturbative results from the statistical atom are of amazing accuracy, but because our focus is on extended systems, self-consistent solutions of atoms are more interesting to us. We therefore explore a self-consistent solution to Englert and Schwinger’s model and assess its quality by comparing it to two other representative orbital-free models that belong to the Thomas-Fermi-Dirac (TFD)-$\lambda$vW family of orbital-free models. Results show that the ES model generally offers the same accuracy as the well-known TFD-$\frac{1}{5}$vW model, but it works better for few quantities, especially for the Pauli potential. We also discerned that the model’s current limitation is the inability to model atoms with a low atomic number (<12).

\section{The Model}
We briefly present the potential functional formalism in DFT and show how it can be used to correct the Thomas-Fermi description of atoms.
\subsection{Formalism}
We start with the usual DFT formalism, where energy is separated as 
\begin{align}
\label{eq:dft_functional}
	E[n] = E_{\text{kin}}[n] + E_\text{int}[n] + \int d\rv V_\text{ext}(\rv)n(\rv),
\end{align}
where $E_\text{int}$ is the functional containing electron–electron interaction. We enforce the particle number $N$ restriction via Lagrange multiplier $\zeta$ and apply the Legendre transformation to the kinetic energy
\begin{align*}
E_1[V + \zeta] = E_\text{kin}[n] + \int \dr \left( V(\rv) + \zeta \right)n(\rv),
\end{align*}
where we introduced an effective single particle potential $V$.

By incorporating this into functional \eqref{eq:dft_functional}, we arrive at a joint functional where the variables $n, V$ and $\zeta$ are treated as independent variables \cite{leading_gradient_correction_2d}
\begin{align*}
E[V,n,\zeta] = E_1[V + \zeta] 
- \int (d\rv)\left[ V(\rv) - V_\text{ext} \right]n(\rv) 
+ E_\text{int}[n] - \zeta N.
\end{align*}
The variations of this functional lead us to the relations
\begin{align}
\label{eq:variation_n}
n(\rv) &= \frac{\delta}{\delta V}E_1[V + \zeta], \\
\label{eq:variation_N}
N &= \frac{\delta}{\delta \zeta} E_1[V + \zeta] = \int \dr n(\rv), \\
\label{eq:variation_V}
V &= V_\text{ext} + \frac{\delta}{\delta n} E_\text{int}[n] = V_\text{ext} + V_\text{int},
\end{align}
where $V_\text{int}$ is the single particle interaction potential. Here we use the binding energy $\zeta$ instead of the more common chemical potential $\mu = -\zeta$. The energy $E_1[V + \zeta]$ is possible to approximate semiclassically \cite{1Airy-averaged_corrections_2d,semiclassical_atom}.

We emphasize here that working with this joint functional is still working within the DFT framework given by the Hohenberg-Kohn theorems, although in this formalism we acknowledge effective potential and chemical potential as independent variables. For simple cases like the Thomas-Fermi theory, it is possible to reduce the potential functional version of $E_1[V + \zeta]$ to density functionals and vice versa \cite{semiclassical_atom}. Because we are working with a single particle potential and $E_1$ will be approximated with the help of a single particle potential, the kinetic energy in this model is non-interacting kinetic energy, which Kohn-Sham and other OFDFT schemes also use.

\subsection{Semiclassical $E_1$}
Englert and Schwinger derived a semiclassical Airy-average expression for $E_1$; this can be understood as a Thomas-Fermi expression that contains quantum corrections. The semiclassical expression is given in terms of Airy-average functions $F_m$, which are closely related to Airy functions.

The derivation is based on the approximation that each electron moves in a local harmonic potential. Then this is expanded in terms of $\rv, \mathbf{p}$ commutators to a first order (Thomas-Fermi corresponds to a zeroth-order approximation; i.e., the position and momentum commute). \cite{semiclassical_atom}. 

The main quantity in the quantum corrected theory is 
\begin{align*}
y = 2(V + \zeta)|2\nabla V|^{-2/3},
\end{align*}
and the Airy-average functions are functions of this variable
\begin{align*}
F_0(y) &= \left[\Ai(y)\right]^2, \\
F_1(y) &= -y\left[\Ai(y)\right]^2 + \left[\Ai'(y)\right]^2, \\
F_{-1}(y) &= -2\Ai(y)\Ai'(y).
\end{align*}
We can obtain the remaining Airy-average functions recursively with the rule
\begin{align*}
(m - \frac{1}{2})F_m(y) = \frac{1}{4}F_{m-3}(y) - yF_{m-1}(y).
\end{align*}

The quantum-corrected Thomas-Fermi expression for $E_1[V + \zeta]$ in terms of Airy-average functions is
\begin{align}
\label{eq:es_energy}
E_1[V + \zeta] = -\frac{1}{4\pi}\int \dr\left[ |2\nabla V|^{5/3}F_3(y) -  \frac{1}{3}\nabla^2 V |2\nabla V|^{1/3} F_1(y) \right],
\end{align}
with approximations to the second order in potential $O(\nabla^2 V)$. Reducing this to a kinetic energy functional \cite{semiclassical_atom} results in a Thomas-Fermi kinetic energy functional plus one-ninth of the von Weizsäcker kinetic energy functional, which is more widely known as the gradient-corrected Thomas-Fermi.  

\subsection{Partition of $E_1[V + \zeta]$}
We note that the semiclassical evaluation is not valid near the atomic nucleus, which is why it is reasonable to split the evaluation of energy into two parts: Electrons described well by the Thomas-Fermi theory and strongly bound electrons (SBE) that are better described as hydrogenic states \cite{schwinger_leading_correction}. This is the main motivation for the joint functional treatment, as similar treatment is unavailable in terms of density functionals.

Energy $E_1$ is the trace over the non-interacting atomic Hamiltonian
\begin{align*}
E_1[V + \zeta] = \text{Tr}(H + \zeta)\eta(-H-\zeta),
\end{align*}
where the Hamiltonian is $H = \frac{1}{2}\mathbf{p}^2 + V(\rv)$ and $\eta$ is the Heaviside function. We can split this into two parts using an energy scale given by $\zeta_s$, so that the electrons with energy levels above $\zeta_s$ are treated semiclassically and electrons with energy below $\zeta_s$ are treated with discrete quantum states. We accomplish this by adding an intelligent zero based on $\zeta_s$. This results in

\begin{align*}
E_1 = \underbrace{E_1[V + \zeta] - E_1[V + \zeta_s] + (\zeta - \zeta_s)N[V + \zeta_s]}_{E_{\zeta\zeta_s}} + E_S,
\end{align*}
where the functional $E_{\zeta\zeta_s}$ is approximated semiclassically and $E_S$ is evaluated with some other quantum mechanical method. Formally this can be thought of as the evaluation of the semiclassical approximation from which we subtract semiclassical evaluation on energy scales below $\zeta_s$ and finally adding the correct treatment for electrons below energy $\zeta_s$.

The critical part of the ES model is to treat the exact part $E_S$ as hydrogenic states. It is assumed that the potential near the nucleus is quite close to Coulombic potential $-\frac{Z}{r}$ so that perturbative evaluation of the states is accurate enough
\begin{align*}
E_S = Z^2n_s + \int \dr (V + \frac{Z}{r})n_\text{SBE},
\end{align*}
where $n_s$ is the uppermost electron shell treated with hydrogenic states and $n_\text{SBE}$ is the density of the electrons treated with hydrogenic states. By shell, we mean the collection of all hydrogenic states with same energy; thus the shells are tabulated by the quantum number $n$. The density $n_\text{SBE}$ is obtained from spherically averaged hydrogenic states. 
\begin{align*}
n_\text{SBE} = \sum_{i = 1}^{i \leq n_s}|\psi_i|_\text{av}^2,
\end{align*}
where $|\psi_i|_\text{av}$ is the spherically averaged wavefunction of $i$th hydrogenic shell. 
For justification and details, see \cite{schwinger_leading_correction,statistical_atom_1}. Originally, Englert and Schwinger investigated corrections to the Thomas-Fermi model—and as they pointed out, the cut-off energy $\zeta_s$ has certain ambiguity because we are trying to patch a continuous semiclassical model with a discrete quantum model. The most consistent way to achieve this for the Thomas-Fermi model is to average over two electronic shells with equal weights on an energy scale. This is also a good choice for the Thomas-Fermi model with corrections used here.

More concretely, this means
\begin{align*}
E_{\zeta\zeta_s} \longrightarrow \sum_{i = 1}^2 \frac{1}{2}E_{\zeta\zeta_i},
\end{align*}
where $\zeta_i$ is the corrected binding energy of $i$th electronic shell
\begin{align}
\label{eq:def_zeta_i}
\zeta_i = \frac{Z^2}{2n_i^2} - \int \dr (V + \frac{Z}{r})|\psi_i|_\text{av}^2.
\end{align}
For practical purposes, these corrections can be embedded inside the Airy-average functions $F_m$; thus they become corrected $F_m$. The correction induces their own $y$ variables
\begin{align*}
y_i = 2(V + \zeta_i)|2\nabla V|^{-2/3},
\end{align*}
and by doing replacement
\begin{align*}
F_m \longrightarrow \sum_{i = 1}^2\frac{1}{2}w_j\left[F_m(y) - F_m(y_i) - (y_j - y)F_{m-1} \right],
\end{align*}
we can use the energy expression \eqref{eq:es_energy}. The corrections for strongly bound electrons are inside the functions $F_m$.

Our numerical results indicate that for all atoms of the relevant size ($Z < 100$), the only energetically believable correction is the one where only the lowest shell (i.e., 1s electrons) is treated exactly, and average occurs over shells 1 and 2. This approximation naturally breaks down when the number of total electrons is low, namely $Z \sim 12$.

\subsection{Semiclassical Density}
The density can be calculated via relation $n = \frac{\delta}{\delta V}E_1$ \eqref{eq:variation_n}. The resulting density splits straightforwardly into two contributions
\begin{align*}
n = \frac{\delta}{\delta V}E_1 = \underbrace{\left(\frac{\partial e_1}{\partial V} - \nabla \cdot \frac{\partial e_1}{\partial \nabla V} + \nabla^2 V\frac{\partial e_1}{\partial \nabla^2 V} \right)}_{\tilde{n}} + 
\frac{\delta \zeta_s}{\delta V}\frac{\partial}{\partial \zeta_s}E_1 + n_\text{SBE}
,
\end{align*}

where $e_1$ is the energy density of $E_1$, $\tilde{n}$ is the semiclassical contribution, and 
the two last terms are the contribution of the strongly bound electrons
. Note that the contribution from strongly bound electrons contains the derivative with respect to the core binding energy $\zeta_s$, because it depends on the potential via \eqref{eq:def_zeta_i}. The derivative is
\begin{align}
\label{eq:def_Q}
\frac{\delta \zeta_s}{\delta V}\frac{\partial}{\partial \zeta_s}E_1
= \sum_{i = 1}^2 \frac{1}{2} \underbrace{(\zeta_i - \zeta)\left[\int \dr \frac{1}{\pi} |2\nabla V|^{1/3} F_1(y_j) - \frac{1}{3}|2\nabla V|^{-1}\nabla^2 V F_{-1}(y_j)\right]}_{:=\; Q_i}|\psi_i|_\text{av}^2,
\end{align}
so there will be contributions from the hydrogenic states that depend on the potential.

The semiclassical contribution $\tilde{n}$ in spherical symmetry is
\begin{align}
\label{eq:full_density}
\tilde{n} =& \frac{1}{2\pi}|2\nabla V| F_2 - \frac{1}{6\pi}|2\nabla V|^{1/3}\nabla^2 V F_0 \\
\nonumber
&+ \frac{1}{r}\left[
\frac{1}{36\pi}|2\nabla V|^{2/3}F_{-2} 
- \frac{1}{9\pi}|2\nabla V|^{-2/3}\nabla^2 VF_{-3}
- \frac{1}{108\pi} |2\nabla V|^{-2/3}\nabla^2 VF_{-5}\right] \\
\nonumber
&+ \frac{1}{r^2}\left[ 
\frac{1}{36\pi}|2\nabla V|^{1/3}F_{-2} + \frac{1}{108\pi}|2\nabla V|^{1/3}F_{-5}
\right],
\end{align}
where the first term $\frac{1}{2\pi}|2\nabla V| F_2$ contains the Thomas-Fermi limit. The expression is valid only for neutral atoms and positive ions, as it assumes that the potential’s gradient is positive.
Theoretically, using a simpler density expression is also a viable option, which is only the first two terms of \eqref{eq:full_density}. For our benchmarking, we use only the aforementioned density, but later we discuss the self-consistent accuracy of the simpler density.

As in the case of energy, we can include the correction for strongly bound electrons by using the corrected Airy-average functions $F_m$.

\subsection{Interaction}
\label{sec:interaction}
So far, we have detailed how to calculate $E_1$ and electronic density, which are independent of any possible interaction. Now we detail the electronic interaction included via the effective potential $V$.

The interaction term $E_\text{int}$ is separated into an electrostatic term (Hartree) and into an exchange term
\begin{align*}
E_\text{int} = E_\text{H} + E_\text{ex}.
\end{align*}
The corresponding interaction potential is then
\begin{align*}
V_\text{int} = \frac{\delta E_\text{int}}{\delta n} = V_\text{H} + V_\text{ex}.
\end{align*}
The Hartree term is well known
\begin{align*}
V_\text{H} = \int \dr' \frac{n(\rv)}{|\rv - \rv'|},
\end{align*}
and its connection to density will be used to achieve a self-consistent solution. We do not add correlation effects to the model, as the effect of correlation in atoms is below the accuracy of semiclassical $E_1$ \cite{semiclassical_atom}.

Before the model is complete, we must comment on how to add the exchange effects. The aim is to include local exchange effects, as described by the Dirac exchange functional \cite{lda_dirac}
\begin{align}
\label{eq:dirac_ex}
E_\text{ex} = - \frac{1}{4\pi^3}\int \dr (3\pi^2n(\rv))^{4/3}.
\end{align}

To be consistent, we discard exchange effects on strongly bound electrons, because we already approximated them to be non-interacting (although they will receive a marginal contribution via potential). The exchange potential is necessary for self-consistency, given by 
\begin{align}
\label{eq:V_ex_n}
V_\text{ex} = \frac{\delta}{\delta n}E_\text{ex} = -\frac{1}{\pi}(3\pi^2n(\rv))^{1/3}.
\end{align}
The most straightforward way to add exchange effects is simply by replacing $n \rightarrow \tilde{n}$ so that only the smooth part is treated with exchange. Testing showed us that while this method produced exchange effects of the correct magnitude, it is not the most accurate way to include the exchange (while discarding contributions of the strongly bound electrons).

The second route taken by Englert and Schwinger is to start from the Thomas-Fermi theory to arrive at a potential description of the exchange. At the Thomas-Fermi level, the exchange potential can be calculated from $V_\text{ex} = \pi \frac{\partial}{\partial V}n$, and the exchange energy density $\epsilon_\text{ex}$ from relation $\frac{\epsilon_\text{ex}}{\partial y} = \frac{1}{2\pi}|2\nabla V|^{4/3}V_\text{ex}^2$. To include exchange effects on the Thomas-Fermi level, the obvious choice for $n$ here is the Thomas-Fermi part of the density $\frac{1}{2\pi}|2\nabla V|F_2$, resulting in the potential
\begin{align}
\label{eq:V_ex_1}
V_\text{ex} = -|2\nabla V|^{1/3}F_1
\end{align} 
and the exchange energy
\begin{align*}
E_\text{ex} = \int \dr \; \frac{1}{2\pi}|2\nabla V|^{4/3}\left( -\frac{1}{4}F_1F_{-1} + \frac{1}{8}F_0^2 + \frac{1}{2}yF_1^2\right).
\end{align*}

We found the description of the exchange potential inadequate at larger distances in the atoms. Thus, we also use another version of the exchange potential \cite{statistical_atom_2}, which uses the density expression $\tilde{n}$ where the Laplacian of the potential has been approximated out. The resulting exchange potential is 
\begin{align}
\label{eq:V_ex_2}
V_\text{ex} = -|2\nabla V|^{1/3}(F_1 - \frac{1}{6}F_{-2}),
\end{align}
which we will use in our implementation. We expect the previous exchange energy description to be accurate enough for this potential. Later we discuss different exchange approximations that are meaningful.

\section{Implementation}

We obtained the self-consistent solution via relation $n = \frac{\delta}{\delta V}E_1$ and the connection of a single particle potential $V$ to the electrostatic potential $V_H$. For a given effective potential $V$, we find the corresponding binding energy $\zeta$ with relation $N = \int \dr n[V(\rv)] + \zeta]$. This yields the density $n[V(\rv) + \zeta]$, which we can use to find a new Hartree potential $V_H$ through a Poisson equation. The new effective potential is now obtained from this Hartree potential with $V = V_H + V_\text{ext} + V_\text{ex}[n, V]$, where in $V_\text{ex}$ we use the old potential and density. This is iterated until the change in potential and binding energy is small enough. With a good initial guess, this procedure gives the self-consistent solution. Here the Thomas-Fermi potential is a sufficient initial guess. 

We describe this idea in a bit more detail for spherically symmetrical atoms. With relation \eqref{eq:variation_V} and the connection of an electrostatic potential $V_H$ with a Poisson equation, we arrive at
\begin{align*}
-\frac{1}{4\pi}\nabla^2\left(V + \frac{Z}{r}\right) = n(\rv) - \nabla^2 V_\text{ex},
\end{align*}
with boundary conditions $rV(\rv \longrightarrow 0) \longrightarrow -Z$  and $rV(r \longrightarrow \infty) = -Z + N$, where zero potential has been assigned to infinitely far away from the nucleus. After we use the fact that our system is spherically symmetric and introduce the auxiliary quantity $V(\rv) = -\frac{Z}{r}\Phi(r)$, we have the differential equation
\begin{align}
\label{eq:differential_eq}
Z\frac{\partial^2}{\partial r^2}\Phi(r) &= 4\pi rn(r) - \frac{\partial^2}{\partial r^2}rV_\text{ex},\quad \text{with boundary conditions} \\
\nonumber
\Phi(0) &= 1, \\
\nonumber
\Phi(\infty) &= 1 - \frac{N}{Z}.
\end{align}

After obtaining the effective potential $V$, we find the corresponding binding energy with Newton’s method from the relation $N -\int \dr n(V + \zeta) = 0$. 
We use a non-uniform grid that is denser near the nucleus, where the change in values is greater than in the tail (the grid is correspondingly sparser in the tail).

\subsection{Numerical Method}
Englert and Schwinger’s original paper \cite{statistical_atom_3} uses the shooting method to solve the resulting differential equation. We solve the resulting differential with a simple 1D finite element method with linear elements. For the finite element method, we derive the weak form of the differential equation, which is
\begin{align*}
-Z\int dr \; \frac{\partial \Phi(r)}{\partial r}\cdot\frac{\partial v}{\partial r}
= \int dr \; 4\pi rn(r) \cdot v + \frac{\partial\; rV_\text{ex}(r)}{\partial r}\cdot\frac{\partial v}{\partial r},
\end{align*}
where $v$ is the test function. During testing we noted that the exchange potential is the most sensitive quantity from a numerical point of view. In the numerical study \cite{statistical_atom_3}, the solution for neutral atoms was not obtained because the effective potential’s long-range behavior is unknown. 

We obtain the neutral atom solution simply by setting the boundary condition to zero, as dictated by \eqref{eq:differential_eq}, and then we converge the result with respect to grid size until the errors (due to the grid’s finite size) are below the error threshold. 

The differential equation is a bit different than Englert and Schwinger’s book \cite{semiclassical_atom}. We use the more straightforward expression derived from a Poisson equation than they mention in the book \cite{semiclassical_atom}, where all the terms containing the Laplacian of effective potential are moved to the left side of the equation.

\section{Results}
We benchmark the numerical results against perturbative results calculated by Englert and Schwinger, the Kohn-Sham results, and the TFD-$\lambda$W model. First we evaluate our results against Englert and Schwinger’s \cite{semiclassical_atom}, and then compare the self-consistent results for the following methods:
\begin{itemize}
\item[ES] The Englert-Schwinger model with exchange potential \eqref{eq:V_ex_2} and density expression \eqref{eq:full_density}.
\item[KS-LDA] Spherically symmetric Kohn-Sham atom, where the xc-functional is the Dirac exchange \eqref{eq:dirac_ex} and no correlation (KS-LDA stands for Kohn-Sham local density approximation).
\item[TFD-$\frac{1}{9}$W] The OFDFT model, which contains Dirac exchange and Thomas-Fermi plus one-ninth of a von Weizsäcker term as a kinetic energy functional, where the von Weizsäcker factor is derived as a quantum correction to the Thomas-Fermi functional. 
\item[TFD-$\frac{1}{5}$W] The OFDFT model, which contains the Dirac exchange and Thomas-Fermi plus one-fifth of a von Weizsäcker term as the kinetic energy functional, where the von Weizsäcker factor is fitted rather than derived.
\end{itemize}
All of these models are based on DFT. In each model, we treat the exchange effect with the Dirac exchange \eqref{eq:dirac_ex}. Thus we choose to benchmark against Kohn-Sham as the most accurate of the approximations.  Originally, Englert and Schwinger opted to benchmark against Hartree-Fock data.

The exact difference between the orbital-free models is a bit more complex. Both orbital-free and ES are (in a sense) approximating the energy $E_1$. The ES model approximates it directly, including both kinetic and potential terms, while TFD-$\lambda$W models approximate it via approximating the kinetic energy density functional only. From a formal point of view TFD-$\frac{1}{9}$W and the ES model both expand the semiclassical trace to the same order in $\hbar$, but we must remember that the TFD-$\frac{1}{9}$W model disregards the strongly bound electron correction completely. In model TFD-$\frac{1}{5}$W, the von Weizsäcker fraction is fitted to produce best results for atoms \cite{tfdvw_atom_fitted_lambda}.

The atomic Kohn-Sham solver used is available in a grid-based implementation of the projector-augmented waves (GPAW) DFT package \cite{gpaw}. 
The kinetic energy density functional methods are implemented within this same Kohn-Sham solver. The implementation details are described elsewhere \cite{ofdft_paw,ofdft_atoms}.

\subsection{Comparison to Englert-Schwinger Reference Numerical Data}
We first study the model presented in the book {\it Semiclassical Theory of Atoms} \cite{semiclassical_atom}. Thus, we use density expression \eqref{eq:full_density} and exchange potential \eqref{eq:V_ex_1} and solve the resulting differential equation. 

We compare the numerical results against the ones provided in \cite{semiclassical_atom} in Table \ref{tab:numerical_ES_comparison} for Krypton electronic configuration. The numbers correspond to fair accuracy. The biggest error is the binding energy of $Z$=38, charge=2 where we have a deviation of 3.3\%. Deviation in other values are below 2\%. Core binding energies $\zeta_1$ and $\zeta_2$ show strong similarity, which is to be expected as the potential should have a $-\frac{Z}{r}$ shape near the hydrogenic states.
The neutral atom results by Englert and Schwinger are the result of extrapolation, so they are omitted from the comparison.

\begin{table}[!tbp]
\begin{tabular}{c |c | l | l | l | l | l | l |}
 Z & charge & Source & $\zeta$ & $\zeta_1$ & $\zeta_2$ & $Q_1$ & $Q_2$\\
\hline 
36 & 0
& This work & 0.0315 & 496.31 & 59.98 & 0.8190 & 4.044 \\
\hline
37& 1
& This work & 0.3311 (1.1) & 526.62 (0.5) & 64.816 (1.9) & 0.8225 (0.3) & 4.0824 (1.3) \\
& & Ref \cite{semiclassical_atom} & 0.3347 & 529.42 & 66.103 & 0.8251 & 4.137 \\
\hline
38 & 2
&This work & 0.7432 (3.3)  & 558.96 (0.4)  & 69.703 (0.2) & 0.8244 (0.4) & 4.127 (1.0) \\
& & Ref \cite{semiclassical_atom} & 0.7687 & 561.29 & 71.377 & 0.8276 & 4.168 \\
\end{tabular}
\caption{Comparison of numerical parameters. Error compared to reference values \cite{statistical_atom_3} are in parenthesis (in percent).
}
\label{tab:numerical_ES_comparison}
\end{table}

\subsection{Effect of Exchange Potential}

Exchange potential and energy have few possible approximations, as indicated in section \ref{sec:interaction}. To choose the best one, we take a look at Krypton to see the effects of different exchange functionals. We look at the total energy, binding energy $\zeta$, and averages over densities $\langle \frac{1}{r} \rangle$, $\langle r \rangle$ and  $\overline{r}^2$. The last one is defined by
\begin{align*}
\overline{r}^2 = \frac{1}{N}\int \dr \; r^2n(\rv).
\end{align*}
The results for different exchange expressions for neutral Krypton are tabulated in Table \ref{tab:ex_potentials}.

From the change $\overline{r}^2$ we can see that the choice of exchange potential has a strong effect near the atom’s edge, which is quite natural \cite{semiclassical_atom}.
As the exchange energy is negative and the corresponding potential is attractive, we would expect the correct inclusion of exchange to make the atomic size smaller. 

The energy difference is not that useful on a semiclassical scale if we compare it to a highly accurate Englert-Schwinger prediction for semiclassical energy, which is $-2747.64$ Hartrees for Krypton. The deviation from this value is 0.7\%, 0.1\% and 0.2 \% for methods \eqref{eq:V_ex_n}, \eqref{eq:V_ex_1}, and \eqref{eq:V_ex_2}, respectively. Our choice, then, should be based on the quality of density at the atom’s outer reaches. The experimental value for $\overline{r}^2$ can be obtained via diamagnetic susceptibilities \cite{statistical_atom_3}, which are provided by the {\it CRC Handbook of Chemistry and Physics} \cite{experimental_data}. The experimental value of $\overline{r}^2$ for Krypton is $1.010$ Bohr$^2$. This indicates that all methods are a bit insufficient, but that \eqref{eq:V_ex_2} is clearly the best for $\overline{r}^2$. Finally, we want to note that there is a deviation of 0.05 - 0.1 Bohr$^2$ to the reported $\overline{r}^2$values\cite{statistical_atom_3} so that the deviation to the experimental value might already be at the level of precision of the chosen methods.

\begin{table}[!tbp]
\begin{tabular}{l | c | c | c | c | c | c |}
Exchange potential $V_\text{ex}$ & $E$ & $\zeta$ & $\langle \frac{1}{r} \rangle$ & $\langle r \rangle$ & $\overline{r}^2$ & $r_\text{classical}$ \\
\hline
\eqref{eq:V_ex_n} $-\frac{1}{\pi}(3\pi^2\tilde{n}(\rv))^{1/3}$  & -2765.852 & 0.08867 & 186.618 & 29.0069 & 1.3718 & 4.7613 \\
\eqref{eq:V_ex_1} $-|2\nabla V|^{1/3}F_1$ & -2744.064 & 0.03152 & 187.351 & 28.6654 & 1.3650 & 4.9333 \\
\eqref{eq:V_ex_2} $-|2\nabla V|^{1/3}(F_1 - \frac{1}{6}F_{-2})$ & -2742.266 & 0.02609 & 187.744 & 27.3852 & 1.1655 & 4.1455
\end{tabular}
\caption{Krypton $Z=36$ with different exchange potentials with full density expression \eqref{eq:full_density}. Classical radius is defined by $V(r_\text{classical}) + \zeta = 0$. }
\label{tab:ex_potentials}
\end{table}

We therefore determined that for the self-consistent ES atom model, the exchange potential \eqref{eq:V_ex_2} is preferred. Still, the model’s accuracy in the atom’s outer reaches is somewhat limited by the exchange approximation, and it remains unclear how the exchange effect should be treated for strongly bound electrons. We intend to address this point in future work. 

In the following, we use exchange potential \eqref{eq:V_ex_2}.

\subsection{Assessment of ES Improvements in Neutral Atoms}
We compute the energies and a few other descriptive quantities for the ES model and compare them to other models for neutral systems to assess the quality of the self-consistent solution. The total energy as a quantity is not so important, as the real predictive power lies in the energy differences, but having this information is somewhat useful, because it informs the quality of approximations made. As previously mentioned, though, here we consider Kohn-Sham as the “ground truth.”

Most of our quantities depend on density to indicate the shape and quality. The near-nucleus area is probed by averaging $\langle 1/r \rangle$, which is related to the shielding of the nuclear magnetic moment \cite{semiclassical_atom}. The average over $r$ is related to electric polarizability \cite{semiclassical_atom}. Finally, as we mentioned in the previous section, we measure $\langle r^2 \rangle$, which probes the density at atoms’ outer edges and is related to diamagnetic susceptibility. 

For atoms, the energy difference is tested only by calculating ionization potential and comparing the total energies of charged and neutral systems. As the semiclassical model incorporates no information of the valence electron shells, our primary interest is in the ionization potential of alkali metals, where the semiclassical approximation could produce good results. 

\begin{figure}[ !tbp]
\includegraphics[width=.7\textwidth]{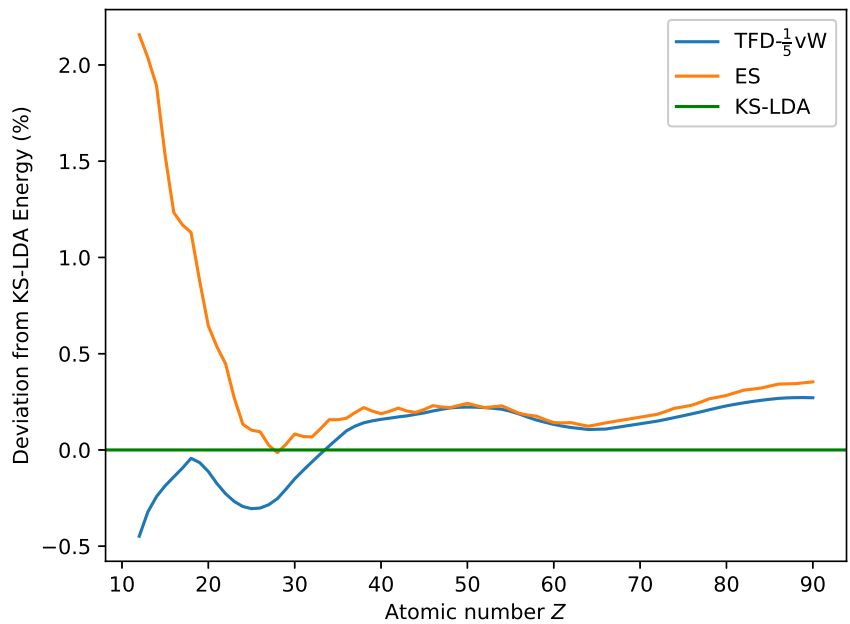}
\caption{Total energy error compared to the Kohn-Sham energy as a function of $Z$.}
\label{fig:total_energy_error}
\end{figure}

First we show the general trends over the whole $Z$ to get a sense of what is or is not a reasonable comparison. Figure \ref{fig:total_energy_error} shows the relative error when compared to Kohn-Sham energies. It is reassuring to see the error is generally small, and it goes down for both methods when $Z$ goes higher, which means that the methods are capturing the essence of the Thomas-Fermi theory. The larger deviation for small-$Z$ for ES can be explained by the strongly bound electron correction, with averaging over shells not as valid for small-$Z$. For high-$Z$, one possible explanation is that we are correcting for a tad too few electrons here, but adding a second shell of strongly bound electrons does not improve the situation here, as then we are overcorrecting. 

\begin{figure}[ !tbp]
\includegraphics[width=.7\textwidth]{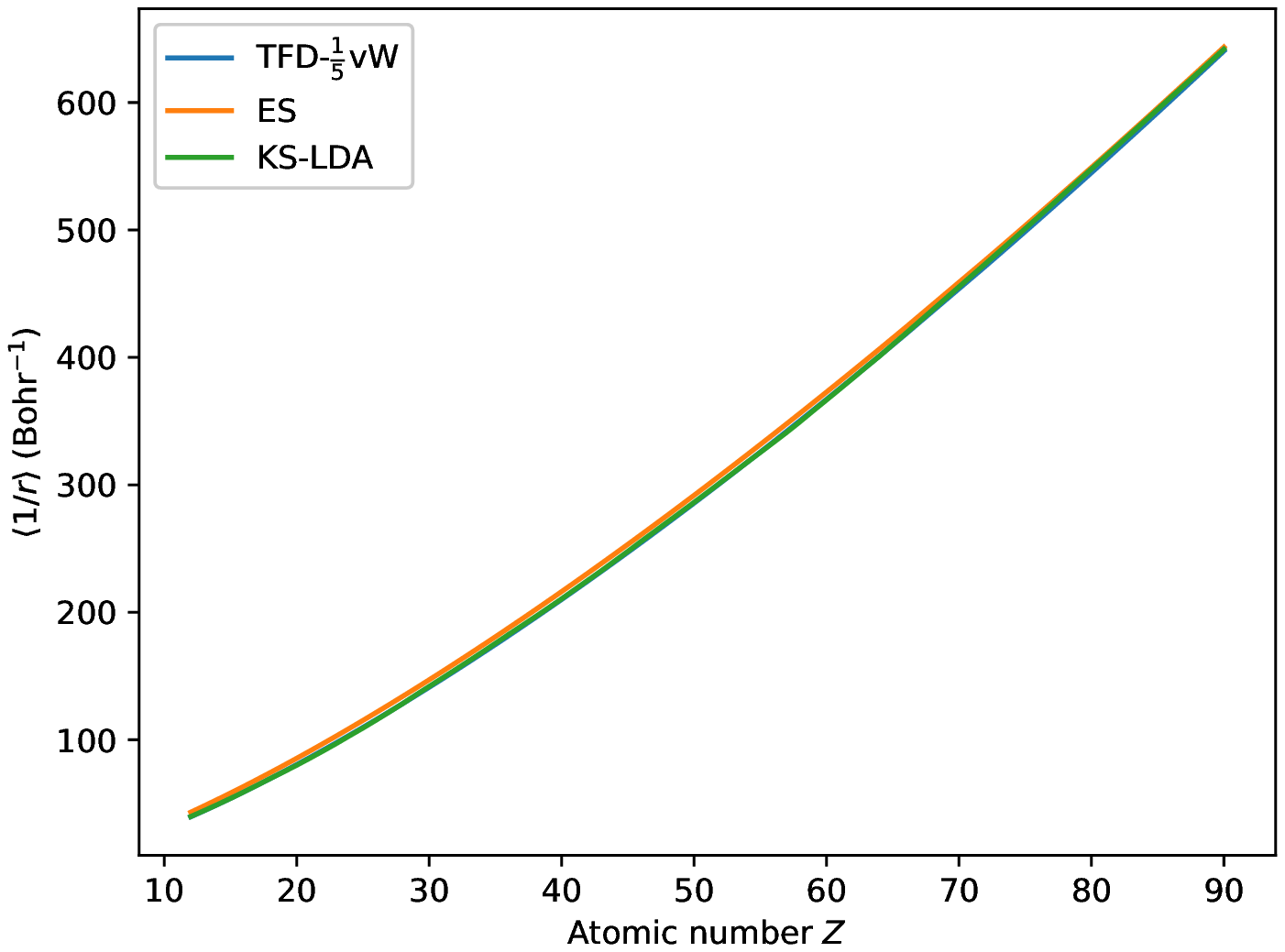}
\caption{Expectation value $\langle \frac{1}{r}\rangle$ as a function of $Z$.}
\label{fig:invr_Z}
\end{figure}

The expectation value $\langle \frac{1}{r} \rangle$ in Figure \ref{fig:invr_Z} is quite smooth for all models, as mainly strongly bound electrons contribute to the density near the nucleus. The ES model handles them explicitly, while the TFD-$\lambda$vW model seems to handle them implicitly with the von Weizsäcker term.

\begin{figure}[ !tbp]
\includegraphics[width=.7\textwidth]{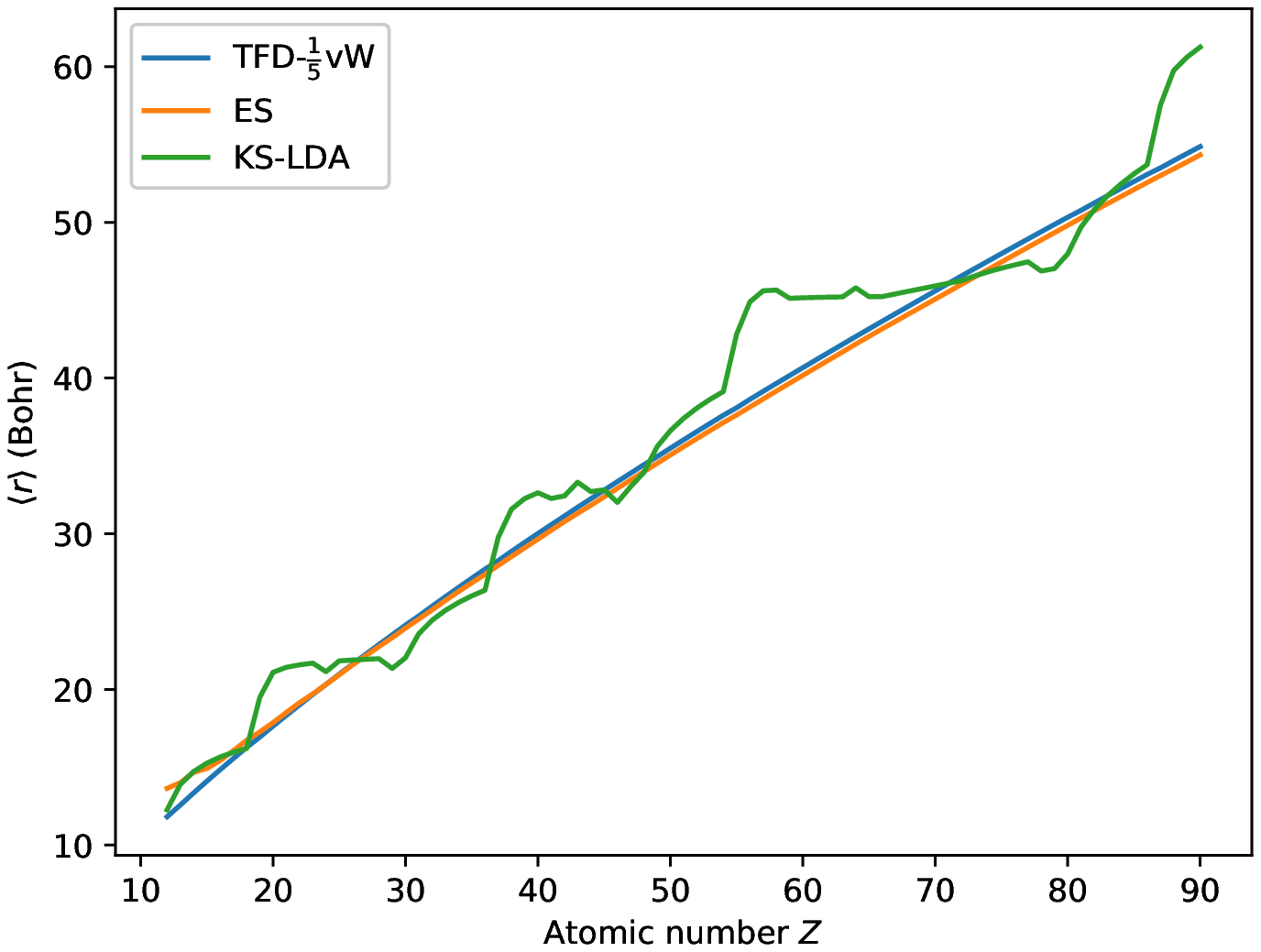}
\caption{Expectation value $\langle r \rangle$ as a function of $Z$.}
\label{fig:r_Z}
\end{figure}

In Figure \ref{fig:r_Z}, we start to see the so-called shell oscillation for $\langle r \rangle$. The main comment here is that we should not take this value too seriously when doing comparisons, as long as the semiclassical model has a decent average over the shell effects. As is obvious, both models satisfy this requirement.

\begin{figure}[ !tbp]
\includegraphics[width=.7\textwidth]{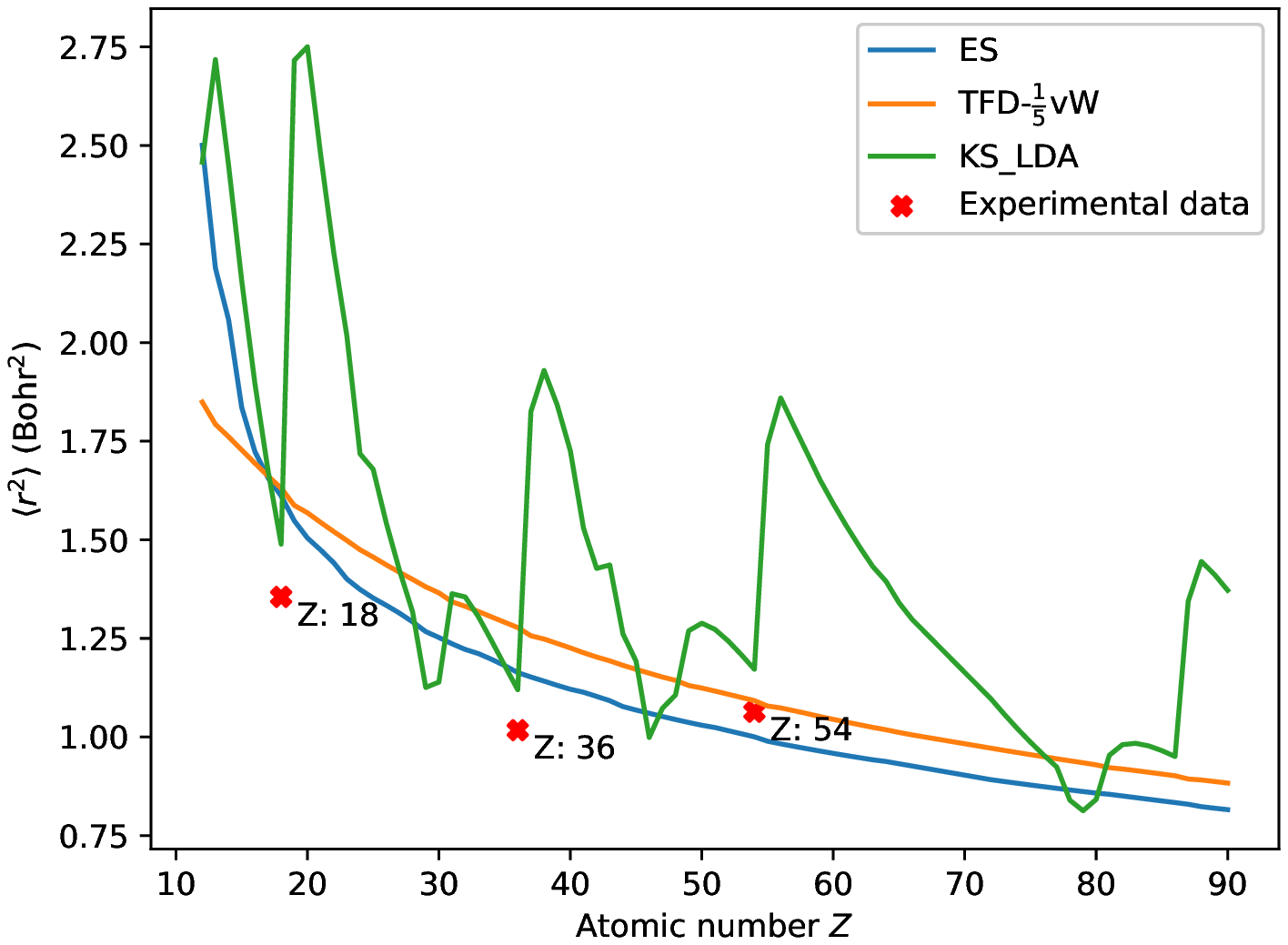}
\caption{Expectation value $\langle r^2 \rangle$ as a function of $Z$. The experimental values are from \cite{experimental_data}.}
\label{fig:r2_Z}
\end{figure}

Next we look at the general trend of $\overline{r}^2$ in Figure \ref{fig:r2_Z}. The shell oscillations already present in the case of $\langle r \rangle$ are magnified, as we are probing even farther reaches of the atoms. We see again that both OFDFT models are reasonable. 

Comparing the ES model with KS-LDA and experimental numbers should only be taken seriously for inert atoms, which have a closed shell structure.
In Xenon ($Z$=54) we see that experimental value is closer to semiclassical value than KS-LDA. The ES model is actually in better agreement with the experimental values as it assigns smaller sizes for almost all atoms\cite{[{See supplemental note}] fn1}.
The effect is mostly due to the exchange term as seen from Table \ref{tab:ex_potentials}.
The origin of these shell oscillations in the Kohn-Sham model is obvious: some outer orbitals are more delocalized than others.

We can see next in the comparison of different atomic models the absolute value of the numbers presented as general trends in Figures \ref{fig:total_energy_error} through \ref{fig:r2_Z}. 
We feel it is informative enough to focus on three representative closed-shell systems: Argon (low-$Z$, where the strongly bound electron approximation is still valid), Krypton (medium-$Z$), and Xenon (high-$Z$). 

\begin{table}[ !tbp]
\centering
\begin{tabular}{l|c|c|c|c|}
Quantity & KS-LDA & TFD+$\frac{1}{9}$vW & TFD+$\frac{1}{5}$vW & ES  \\ 
\hline
E & 								-524.51 & -561.80& -524.75	& -518.60\\
$\langle r \rangle$ & 			16.205 	& 16.202 & 16.267 	& 16.729 		 \\
$\overline{r}^2$ & 				1.489 	& 1.61	 & 1.630		& 1.611	 \\
$\langle \frac{1}{r} \rangle$ & 	69.567 	& 73.796 & 69.566 	& 74.228
\end{tabular}
\caption{Energies and averages of different models for Argon $Z=18$. Results are in atomic units.}
\label{tab:model_quantities_Ar}
\end{table}

\begin{table}[ !tbp]
\centering
\begin{tabular}{l|c|c|c|c|}
Quantity & KS-LDA & TFD+$\frac{1}{9}$vW & TFD+$\frac{1}{5}$vW & ES  \\ 
\hline
E & 								-2746.649 	& -2895.528 	& -2744.153 & -2742.266  \\
$\langle r \rangle$ & 			26.37 		& 27.69 		& 27.72 		& 27.81  \\
$\overline{r}^2$ & 				1.120 		& 1.274 		& 1.277  	& 1.166 \\
$\langle \frac{1}{r} \rangle$ & 182.63 		& 190.14 	& 181.63 	& 186.34 
\end{tabular}
\caption{Energies and averages of different models for Krypton $Z=36$. Results are in atomic units.}
\label{tab:model_quantities_Kr}
\end{table}

\begin{table}[ !tbp]
\centering
\begin{tabular}{l|c|c|c|c|}
Quantity & KS-LDA & TFD+$\frac{1}{9}$vW & TFD+$\frac{1}{5}$vW & ES  \\ 
\hline
E & 								-7223.567 	& -7556.458 & -7208.302 	& -7207.0235 \\
$\langle r \rangle$ & 			39.131 		& 37.610 	& 37.606 	& 38.650  \\
$\overline{r}^2$ & 				1.172 		& 1.092	 	& 1.092 		& 1.000 \\
$\langle \frac{1}{r} \rangle$ & 	317.670 		& 330.690 	& 317.60 	&  323.931
\end{tabular}
\caption{Energies and averages of different models for Xenon $Z=54$. Results are in atomic units.}
\label{tab:model_quantities_Xe}
\end{table}

\begin{table}[ !tbp]
\begin{tabular}{c | c | c}
Atom & ES & KS \\ \hline
K & 0.1310 & 0.1364 \\
Rb & 0.1202 & 0.1319 \\
Cs & 0.1133 & 0.1223
\end{tabular}
\caption{Ionization potential in Hartrees calculated by the energy difference  $E(Z) -E(Z-1)$  for ES and KS models. KS results are calculated with GPAW DFT code\cite{gpaw}, where we used Dirac exchange functional.}
\label{tab:ionization_potential}
\end{table}

From Figures \ref{fig:total_energy_error} through \ref{fig:r2_Z} and Tables \ref{tab:model_quantities_Ar} through \ref{tab:model_quantities_Xe}, we can draw some conclusions. First is the remarkable and well-known accuracy of the TFD+$\frac{1}{5}$vW model and the fact that ES is much better than TFD+$\frac{1}{9}$vW, which has a functional that is formally expanded to the same order in $\hbar$, but is missing the corrections for strongly bound electrons. We did not include TFD+$\frac{1}{9}$vW in Figures \ref{fig:total_energy_error} through \ref{fig:r2_Z}, because the accuracy is substantially lower than for other models. The deviation of TFD+$\frac{1}{9}$vW from KS-LDA for Argon, Krypton, and Xenon is 7.1 \%, 5.4 \%, 4.6 \%, which is substantially higher than for TFD+$\frac{1}{5}$vW or ES, as seen from Figure \ref{fig:total_energy_error}. Overall, it seems that after $Z > 40$, TFD+$\frac{1}{5}$vW and the ES model offer relatively similar accuracy, with TFD+$\frac{1}{5}$vW being slightly better.

\begin{figure}[ !tbp]
\includegraphics[width=.8\textwidth]{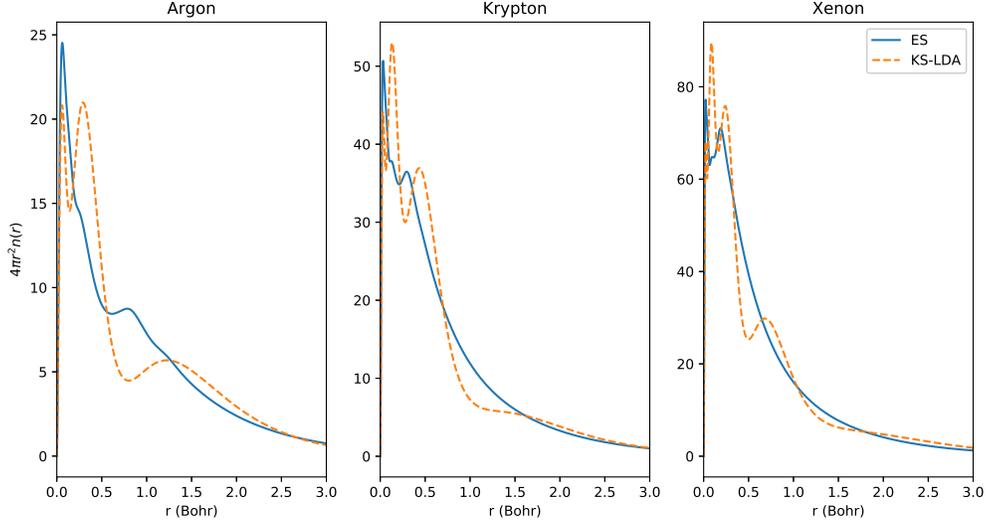}
\caption{Semiclassical and Kohn-Sham densities of inert atoms.}
\label{fig:densities}
\end{figure}

For all models, the density-dependent quantities $\langle r \rangle$, $\overline{r}^2$ and $\langle \frac{1}{r} \rangle$ are quite reasonable, surprisingly even for TFD+$\frac{1}{9}$vW. This reflects the fact that all the models have a reasonable density average over the shell effects in KS-LDA densities, which are shown in Figure \ref{fig:densities}. 

Near the nucleus, TFD+$\frac{1}{5}$vW has the best average description compared to KS-LDA if we look at quantity $\langle \frac{1}{r} \rangle$. The worst deviation for TFD+$\frac{1}{5}$vW is for Krypton with 0.6 \% from KS-LDA, while for TFD+$\frac{1}{9}$vW and for ES the worst case is Argon, with deviations of 6.1 \% and 6.7 \% respectively.

The first quantity to contain shell effects is $\langle r \rangle$. For these values, all the models give surprisingly similar results. The worst case for ES is Argon, where we have a deviation of 3.2 \%, while the worst case for kinetic energy functionals is Krypton, where the deviation is $\sim 5.1 \%$. Again we note that shell oscillation plays a role here and that the Argon value for ES is probably particularly bad: in Figure \ref{fig:r_Z} we see a bump in the ES values for small $Z$. This is most likely a slight artifact caused by the averaging procedure.

The quantity $\overline{r}^2$ is an interesting one. For Argon, all the semiclassical models give similar results, which are significantly above the KS-LDA, up to $\sim 10 \%$ deviation for TFD+$\frac{1}{5}$vW. The case for Krypton is similar, except for ES, which is closer to KS-LDA than kinetic energy functionals. Finally, for Krypton, ES is below all the other models, which is a good thing if we look at the experimental values in Figure \ref{fig:r2_Z}.

For inert atoms there is a trend: all semiclassical values start above KS-LDA values in Argon and end up below KS-LDA in Xenon. However, we should not read too much into this, as the quantity is quite dependent on the outer orbitals of the particular element, as Figure \ref{fig:r2_Z} shows. The ES model clearly wins in the description of the atom’s outer reaches when considering experimental values.

The ionization potential calculated with a difference of total energies $E(N) - E(N-1)$ and the results are in Table \ref{tab:ionization_potential}. We only calculated it for alkali metals, as the ES model will fail to provide reasonable potential for other elements because the shell effects are missing. 
A deviation in the ionization potential emerges (up to 9 \% deviation for Cesium), but the trend is similar for both methods.

Atomic densities for closed-shell atoms are shown in \ref{fig:densities}. We also plot Kohn-Sham densities as a reference. We can see that the ES densities are not completely structureless; they do contain some structure due to averaging over the hydrogenic shells. Yet, obviously, they do not contain the shell structure of Kohn-Sham due to single particle states.

Previous results prove that ES model does not lose to the density functional models and is even better than Kohn-Sham in some special cases.  
But what the ES model mostly supplies is theoretical clarity and rigor. One thorny issue with TFD+$\lambda$vW models and other GGA-based OFDFT models has been the Pauli potential $v_\Theta$’s negativity, which is defined as. 
 \begin{align*}
v_\Theta = \frac{\delta T_\Theta}{\delta n} = \frac{\delta T_s}{\delta n} - \frac{\delta T_\text{vW}}{\delta n} = V - \mu - \frac{\delta T_\text{vW}}{\delta n},
\end{align*}
where $T_s$ is the non-interacting kinetic energy and $T_\text{vW}[n]$ is the von Weizsäcker kinetic energy functional. 
It has been shown that $v_\Theta$ should always be positive \cite{pauli_potential_properties}, but for TFD+$\lambda$vW and many other GGA kinetic functionals, it has been found to be negative near a nucleus \cite{pauli_potential_functionals}. This emerge because the semiclassical evaluation is not valid near a nucleus; see \cite{semiclassical_atom} for an excellent discussion. The ES model does not have this problem, because the problematic region is removed from the semiclassical evaluation.
\begin{figure}[ !tbp]
\includegraphics[width=.8\textwidth]{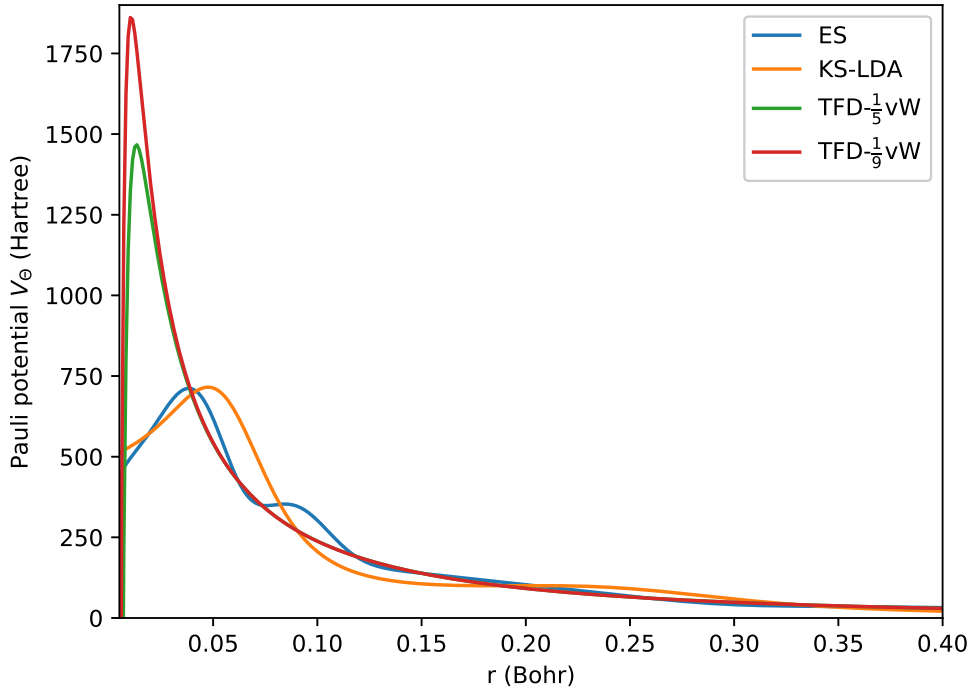}
\caption{Pauli potential $v_\Theta$ of four atomic models.}
\label{fig:pauli_potential}
\end{figure}
Figure \ref{fig:pauli_potential} shows the self-consistent Pauli potentials. Clearly in the self-consistent calculations, the Pauli non-positivity problem is contained in a small region, but the Pauli potential is still qualitatively wrong. In comparison, the ES model’s Pauli potential is qualitatively better, although naturally missing the shell effects. The ES model has oscillations due to the averaging over shells, but this is more of an artifact than a result of the oscillations being in the wrong region.

This raises the question of how the TFD+$\frac{1}{5}$vW’s energetics can be so good (while having reasonable geometric properties, too) if it completely ignores the correction for strongly bound electrons? The answer must be related to the dual nature of the von Weizsäcker term, because it has two roles as a density functional: it is a gradient correction to Thomas-Fermi, but it also is exact for up to two non-interacting particles. These two roles naturally differ by a constant factor, but the form is the same. 
Kinetic energy functionals of GGA type can be built around the dual nature by interpolating between these two extremes \cite{pauli_potential_functionals,ofdft_vt84f,kinetic_energy_perdew}. Here we are inclined to support the view of Englert and Schwinger: it is more straightforward to exclude the strongly bound electrons from the semiclassical evaluation than try to modify the functional to include them via the von Weizsäcker term. The near nucleus Pauli potential singularity is also avoided by the use of pseudopotentials as the strongly bound electrons are excluded from the OFDFT calculation. The non-negativity constraint might be a non-issue for valence density\cite{[{Authors are not aware of any kinetic energy functional that breaks this condition for valence densities}]fn2}.
We also must remember that the $\lambda$ value of $\frac{1}{5}$ is obtained by fitting, while the strongly bound electron correction is better motivated. If the ES model is compared to the gradient expansion in atoms, then it is clearly preferable.

\subsection{Accuracy of Self-Consistent Density Approximations}

As mentioned earlier, it is not necessary to calculate the density with the full expression \eqref{eq:full_density} to get the same semiclassical accuracy. Only the first line of \eqref{eq:full_density} is necessary, as Englert argued \cite{semiclassical_atom}. The rest of the expression contains full divergence—i.e., it integrates to zero so it does not contribute to the number of electrons, only to the distribution. We are interested in the approximation for future use, if the method is extended to larger systems. We also want to see the effect on $\overline{r}^2$ to see how sensitive it is to density approximations, because we already established the effect of exchange.

The simple density approximation is obtained simply by discarding the higher-order variations. The resulting expression is
\begin{align}
\label{eq:simple_density}
\tilde{n}_\text{approx} \approx \frac{\partial e_1}{\partial V} = \frac{1}{2\pi}|2\nabla V| F_2 - \frac{1}{6\pi}|2\nabla V|^{-1/3}\nabla^2 V F_0.
\end{align}

We compare the difference of two density expressions \eqref{eq:simple_density} and \eqref{eq:full_density} in self-consistent calculations. We focus on Krypton for simplicity, but the trends are reproduced over a range of $Z$. As both integrate to the electron number, the difference is purely just a redistribution of the density.

\begin{table}[!tbp]
\begin{tabular}{c | c | c | c | c}
Atom & Density & Energy & $\overline{r}^2$ & $\langle \frac{1}{r} \rangle$ \\
\hline
Ar & $\tilde{n}$ \eqref{eq:full_density} & -518.591 & 1.611 & 74.228 \\
Ar & $\tilde{n}_\text{approx}$ \eqref{eq:simple_density} & -519.082 & 1.652 & 73.301 \\
\hline
Kr & $\tilde{n}$ \eqref{eq:full_density} & -2742.320 & 1.163 & 187.744 \\
Kr & $\tilde{n}_\text{approx}$ \eqref{eq:simple_density} & -2743.970 & 1.193 & 186.344 \\
\hline
Xe & $\tilde{n}$ \eqref{eq:full_density}& -7207.024 & 1.000 & 323.598 \\
Xe & $\tilde{n}_\text{approx}$ \eqref{eq:simple_density} & -7209.345 & 1.014 & 322.737 \\
\end{tabular}
\caption{Results for inert atoms with density expressions \eqref{eq:simple_density} and \eqref{eq:full_density}.}
\label{tab:density_comparison}
\end{table}

From Table \ref{tab:density_comparison} we can see that the simple density assigns more density near the nucleus while the full expression localizes the density more, which is seen in $\overline{r}^2$. Full density has a bit better energy when compared to KS-LDA . For energy, the maximal deviation between density expressions is 0.4 \% percent in Xenon, and for $\overline{r}^2$ the maximal deviation is 3\% percent in Argon. On a semiclassical scale, these differences are quite small.  Overall the differences show that the full density \eqref{eq:full_density} is better, as expected, but the simpler expression \eqref{eq:simple_density} is still valid.

\section{Conclusion}
Having benchmarked the self-consistent ES model, we found that it compares favorably against the simple TFD-$\lambda$vW model. While the self-consistent model does not achieve the accuracy of Englert and Schwinger’s perturbative work, still yet the performance is good and could lead to a viable means of obtaining physically plausible full solutions to OFDFT equations. An important point to remember here is that self-consistent atoms are not valuable themselves, as they are just a stepping stone toward larger systems.
The problematic points of the self-consistent ES model are that it cannot be applied to light elements (according to our tests, the strongly bound electron approximation breaks down numerically at $Z  \sim 12$) and its non-ambiguous inclusion of exchange effects. Given this model’s potential, however, such issues are worth considering and investigating with rigor; these issues can and will be addressed in future work.

\newpage
\bibliography{bibliography}
\end{document}